% version of AAPT talk 2004/01/27 for www.arxiv.org

\magnification=1200
\input eplain
%%%%%%%%%%%%%%%%%%%%%%%%%%%%%%%%%%%%%%%%%%%%%%%%%%%%%%%%%%%%%%%%%%%%%%%%%%%%%%%%
% This file sets up a twelve point environment for TeX. It can be initialized
% with the '\twelvepoint' macro.
% It also sets up a '\tenpoint' macro in case you want to go back down.
% By Don Hosek
%%%%%%%%%%%%%%%%%%%%%%%%%%%%%%%%%%%%%%%%%%%%%%%%%%%%%%%%%%%%%%%%%%%%%%%%%%%%%%%%
 
\ifx\tenpoint\undefined\let\loadedfrommacro=Y
         %%%
%%% Set up the tenpoint macros
%%%
%%% Unhappy things will happen if this file is loaded twice
%%% So don't... as a matter of fact, don't load it directly, let one of the
%%% other size changing macros do it for you.
%%%
%%% Written by Don Hosek
%%%
\ifx\loadedfrommacro Y\else
         \message{10point.TeX must be loaded from a macro package.}
         \message{Input terminated.}
          \fi
 
\font\tencsc=cmcsc10
 
\newfam\scfam
 
\def\tenpoint{\def\rm{\fam0\tenrm}% switch to 10-point type
    \textfont0=\tenrm  \scriptfont0=\sevenrm  \scriptscriptfont0=\fiverm
    \textfont1=\teni   \scriptfont1=\seveni   \scriptscriptfont1=\fivei
    \textfont2=\tensy  \scriptfont2=\sevensy  \scriptscriptfont2=\fivesy
    \textfont3=\tenex  \scriptfont3=\tenex    \scriptscriptfont3=\tenex
    \textfont\itfam=\tenit   \def\it{\fam\itfam\tenit}%
    \textfont\slfam=\tensl   \def\sl{\fam\slfam\tensl}%
    \textfont\ttfam=\tentt   \def\tt{\fam\ttfam\tentt}%
    \textfont\bffam=\tenbf   \scriptfont\bffam=\sevenbf
    \scriptscriptfont\bffam=\fivebf  \def\bf{\fam\bffam\tenbf}%
    \textfont\scfam=\tencsc  \def\sc{\fam\scfam\tencsc}%
    \normalbaselineskip=12pt
    \setbox\strutbox=\hbox{\vrule height8.5pt depth 3.5pt width0pt}%
    \normalbaselines\rm}

         \let\loadedfrommacro=N\fi
 
%%%
%%% Load in the fonts
%%%
\font\twelverm=cmr12
\font\twelvei=cmmi12
\font\twelvesy=cmsy10 scaled \magstep1
\font\twelveex=cmex10 scaled \magstep1
\font\twelvebf=cmbx12
\font\twelvesl=cmsl12
\font\twelvett=cmtt12
\font\twelveit=cmti12
\font\twelvecsc=cmcsc10 scaled \magstep1
 
\font\ninerm=cmr9            \font\sevenrm=cmr7
\font\ninei=cmmi9            \font\seveni=cmmi7
\font\ninesy=cmsy9           \font\sevensy=cmsy7
\font\ninebf=cmbx9           \font\sevenbf=cmbx7
 
%%%
%%% Set up the twelvepoint macros
%%%
\ifx\twelvepoint\undefined
   \def\twelvepoint{\def\rm{\fam0\twelverm}% switch to 11-point type
       \textfont0=\twelverm \scriptfont0=\ninerm \scriptscriptfont0=\sevenrm
       \textfont1=\twelvei  \scriptfont1=\ninei  \scriptscriptfont1=\seveni
       \textfont2=\twelvesy \scriptfont2=\ninesy \scriptscriptfont2=\sevensy
       \textfont3=\twelveex \scriptfont3=\twelveex\scriptscriptfont3=\twelveex
       \textfont\itfam=\twelveit  \def\it{\fam\itfam\twelveit}%
       \textfont\slfam=\twelvesl  \def\sl{\fam\slfam\twelvesl}%
       \textfont\ttfam=\twelvett  \def\tt{\fam\ttfam\twelvett}%
       \textfont\bffam=\twelvebf  \scriptfont\bffam=\ninebf
        \scriptscriptfont\bffam=\sevenbf  \def\bf{\fam\bffam\twelvebf}%
       \textfont\scfam=\twelvecsc \def\sc{\fam\scfam\twelvecsc}%
       \normalbaselineskip=14pt
       \setbox\strutbox=\hbox{\vrule height9.5pt depth4.5pt width0pt}%
       \normalbaselines\rm}
   \fi

%%%%%%%%%%%%%%%%%%%%%%%%%%%%%%%%%%%%%%%%%%%%%%%%%%%%%%%%%%%%%%%%%%%%%%%%%%%%%%%%
% This file sets up an nine point environment for TeX. It can be initialized
% with the '\ninepoint' macro.
% It also sets up a '\tenpoint' macro in case you want to go back down.
% By Don Hosek
%%%%%%%%%%%%%%%%%%%%%%%%%%%%%%%%%%%%%%%%%%%%%%%%%%%%%%%%%%%%%%%%%%%%%%%%%%%%%%%%
 
\ifx\tenpoint\undefined\let\loadedfrommacro=Y%
         %%%
%%% Set up the tenpoint macros
%%%
%%% Unhappy things will happen if this file is loaded twice
%%% So don't... as a matter of fact, don't load it directly, let one of the
%%% other size changing macros do it for you.
%%%
%%% Written by Don Hosek
%%%
\ifx\loadedfrommacro Y\else
         \message{10point.TeX must be loaded from a macro package.}
         \message{Input terminated.}
          \fi
 
\font\tencsc=cmcsc10
 
\newfam\scfam
 
\def\tenpoint{\def\rm{\fam0\tenrm}% switch to 10-point type
    \textfont0=\tenrm  \scriptfont0=\sevenrm  \scriptscriptfont0=\fiverm
    \textfont1=\teni   \scriptfont1=\seveni   \scriptscriptfont1=\fivei
    \textfont2=\tensy  \scriptfont2=\sevensy  \scriptscriptfont2=\fivesy
    \textfont3=\tenex  \scriptfont3=\tenex    \scriptscriptfont3=\tenex
    \textfont\itfam=\tenit   \def\it{\fam\itfam\tenit}%
    \textfont\slfam=\tensl   \def\sl{\fam\slfam\tensl}%
    \textfont\ttfam=\tentt   \def\tt{\fam\ttfam\tentt}%
    \textfont\bffam=\tenbf   \scriptfont\bffam=\sevenbf
    \scriptscriptfont\bffam=\fivebf  \def\bf{\fam\bffam\tenbf}%
    \textfont\scfam=\tencsc  \def\sc{\fam\scfam\tencsc}%
    \normalbaselineskip=12pt
    \setbox\strutbox=\hbox{\vrule height8.5pt depth 3.5pt width0pt}%
    \normalbaselines\rm}

         \let\loadedfrommacro=N\fi
%%%
%%% Load in the fonts
%%%
\font\ninerm=cmr9            \font\sixrm=cmr6
\font\ninei=cmmi9            \font\sixi=cmmi6
\font\ninesy=cmsy9           \font\sixsy=cmsy6
\font\ninebf=cmbx9           \font\sixbf=cmbx6
\font\ninesl=cmsl9           \font\ninett=cmtt9      \font\nineit=cmti9
\font\ninecsc=cmcsc10
\font\ninebfit=cmbxti10 at 9pt% to get PS version (which is only in 10pt)
%%%
%%% Set up the ninepoint macros
%%%
\ifx\ninepoint\undefined
   \def\ninepoint{\def\rm{\fam0\ninerm}%switch to a 9-point type
       \textfont0=\ninerm  \scriptfont0=\sixrm  \scriptscriptfont0=\fiverm
       \textfont1=\ninei   \scriptfont1=\sixi   \scriptscriptfont1=\fivei
       \textfont2=\ninesy  \scriptfont2=\sixsy  \scriptscriptfont2=\fivesy
       \textfont3=\tenex   \scriptfont3=\tenex  \scriptscriptfont3=\tenex
       \def\bfit{\ninebfit}%
       \textfont\itfam=\nineit   \def\it{\fam\itfam\nineit}%
       \textfont\slfam=\ninesl   \def\sl{\fam\slfam\ninesl}%
       \textfont\ttfam=\ninett   \def\tt{\fam\ttfam\ninett}%
       \textfont\bffam=\ninebf   \scriptfont\bffam=\sixbf
        \scriptscriptfont\bffam=\fivebf   \def\bf{\fam\bffam\ninebf}%
       \textfont\scfam=\ninecsc  \def\sc{\fam\scfam\ninecsc}%
       \normalbaselineskip=11pt%
       \setbox\strutbox=\hbox{\vrule height8pt depth3pt width0pt}%
       \normalbaselines\rm}%
   \fi

\font\smallcap=cmcsc10

\font\medbf=cmbx12

\newif\ifpdf
\ifx\pdfoutput\undefined
\pdffalse
\else
\pdftrue
\pdfpagewidth=8.5truein%
\pdfpageheight=11truein%
\fi

\newdimen\size
% lengthen page by advancing vsize and then recentering by adjust voffset
\def\lengthen#1{\size=#1\advance\vsize by \size \advance\voffset
by -0.5\size} 
% same as \lengthen, but in the horizontal direction.
\def\widen#1{\size=#1\advance\hsize by \size \advance\hoffset by
-0.5\size}
%\widen{1.3pt}

\def\boxitspace{3pt}
\long\def\hsurround#1#2{\hbox{#2#1#2}}
\long\def\vsurround#1#2{\vtop{\vbox{#2#1}#2}}

\long\def\myboxit#1{\vsurround{\hsurround{\hsurround
{\vsurround{#1}{\vskip\boxitspace}}{\kern\boxitspace}}\vrule}\hrule}

\def\rdelimit{$\rangle$}
\def\ldelimit{$\langle$}
\ifpdf
\input pdfcolor
\input supp-pdf			%from the ConTeXT package
\pdfinfo {/Author (Sanjoy Mahajan)
/Subject (education reform, science, war)
/Keywords (science, value neutral, power, state, mutual aid)
}
\def\MyBlue{\pdfsetcolor{0.94 0.94 0 0}}
\def\myBlack{\Black}
\def\myRed{\Red}
\def\LinkColor{\MyBlue}

\else  % not putting out pdf
\def\myRed{\textRed}
\def\myBlack{\textBlack}
\input colordvi
\fi

% lots of trickery for automatic hotlinks, either in pdftex or normal tex
\def\rawurlaux#1#2%
{\ldefs
\ifpdf
\ifnum\pdftexversion>13%
  \pdfstartlink
\else \pdfannotlink
\fi
 attr {/Border [0 0 0]} user%
     {/Subtype /Link%
      /A <</S /URI /URI (#1)>>}%
\tdefs
\hbox{%
\ldelimit \LinkColor #2%
\Black \rdelimit}%
\pdfendlink
\else %not running pdftex
\ldefs
\tdefs
\hbox{\textMidnightBlue
\ldelimit #2%
\rdelimit
\textBlack}%

\fi
\egroup}

\def\urlaux#1{\hyphenpenalty=-2000\-\rawurlaux{#1}{#1}}
\def\mailurlaux#1{\rawurlaux{mailto:#1}{#1}}

\begingroup
\catcode`_=13 
\catcode`\&=13
\catcode`\$=13
\catcode`-=13
\catcode`/=13
\gdef\fixu{\def_{\leavevmode \kern.06em \vbox{\hrule width.3em}}}
\catcode`.=13
\gdef\tdefs{\fixu
\def&{\char38}%
\def.{\discretionary{\char46}{}{\char46}}%
\def-{\discretionary{\char45}{}{\char45}}%
\def/{\discretionary{\char47}{}{\char47}}%
\def~{\char126}}

\gdef\ldefs{\chardef\_=`\_ \let_=\_%
\let&=\&%
\let$=\$%
\chardef\~=`\~%
\let~=\~%
\chardef\.=`.%
\let.=\.%
\chardef\-=`-%
\let-=\-%
\chardef\/=`/%
\let/=\/%
}

\gdef\rawurl{\bgroup\catcode`~=13 \catcode`_=13 \catcode`\&=13 \catcode`\$=13
\rawurlaux}

\gdef\url{\hskip0pt plus\hsize\penalty0%
\bgroup\catcode`~=13 \catcode`_=13 \catcode`\&=13 \catcode`\$=13%
\catcode`.=13 \catcode`=13 \catcode`/=13%
\urlaux}

\gdef\mailurl{\bgroup\catcode`~=13 \catcode`_=13 \catcode`\&=13 \catcode`\$=13%
\catcode`.=13 \catcode`=13 \catcode`/=13%
\mailurlaux}

\endgroup

\newbox\ftnumbox
\catcode`@=11
\topskip 10\p@ plus10\p@ \r@ggedbottomtrue
\def\footnote#1{\let\@sf\empty % parameter #2 (the text) is read later
  \ifhmode\edef\@sf{\spacefactor\the\spacefactor}\/\fi
  $^{#1}$%
  \@sf\vfootnote{#1.}}
\def\vfootnote#1{\insert\footins\bgroup
  \ninepoint \baselineskip=9pt%
  \global\setbox\ftnumbox=\hbox{#1\enspace}%
  \hangindent=\wd\ftnumbox \hangafter=1%
  \interlinepenalty\interfootnotelinepenalty
  \splittopskip\ht\strutbox % top baseline for broken footnotes
  \splitmaxdepth\dp\strutbox \floatingpenalty\@MM
  \leftskip\z@skip \rightskip\z@skip \spaceskip\z@skip \xspaceskip\z@skip
  \noindent{\copy\ftnumbox}\footstrut\futurelet\next\fo@t}
\def\f@t#1{#1\@foot}
\catcode`@=12 % at signs are no longer letters

\overfullrule=0pt
\newcount\cuenum \cuenum=0
%\cue was to insert notes to speaker to go to next slide
%\def\cue#1{\global\advance\cuenum by 1%
%\vskip3pt\centerline{==== \it Cue slide \the\cuenum ====}\vskip3pt}
\def\cue#1{\relax}

\newcount\ftnum \ftnum=0

\def\mygap{\nobreak \vskip 6pt \vskip-\prevdepth \noindent
\ignorespaces}

\def\ft{\global\advance\ftnum by 1\footnote{\the\ftnum}}

\def\section#1 \par{\bigbreak
\noindent{\bf #1}\mygap}

\def\beginquote{\begingroup\smallskip\advance\leftskip
by 20truept\advance \rightskip by 20truept%
\ninepoint \ignorespaces\noindent}
\def\endquote{\smallskip\endgroup\noindent}
\def\jnl#1{{\it #1}}

\noindent
{\medbf Physics Education Research:\vskip1pt\noindent
Or it's so hard to find good help these
days}\ft{Copyright 2004/05/03 by Sanjoy Mahajan.  
You may
republish the entire text, in any medium for any purpose, provided
that you preserve this notice, do not change
the text, and do not impose
further restrictions.  Email me \mailurl{sanjoy@mrao.cam.ac.uk} for the
current version.}
\medskip
\noindent{\smallcap Sanjoy Mahajan}
{\it\obeylines \parindent=0pt \parskip=0pt
Cavendish Laboratory/Astrophysics
University of Cambridge
Cambridge CB3 0HE
England
}

\smallskip
\noindent[Based on my talk
at the session `Multiple Goals of Physics Education Research',
AAPT Conference, Miami, FL (27 Jan 2004).]

\advance\baselineskip by 0.5truept

\bigskip

\noindent
Bob Ehrlich\ft{`Does It Matter If Physics Educators Have Well-defined
Goals?', AAPT Conference, Winter 2004, online at
\url{http://www.inference.phy.cam.ac.uk/sanjoy/science-society/}}
would like physics education research (PER) to produce more
physics majors, and Karen Cummings\ft{`PER's Bread and Butter',
AAPT Conference, Winter 2004; online at
\url{http://www.inference.phy.cam.ac.uk/sanjoy/science-society/}}
warns PER not to stray far from
problem solving.  But whose problems do we teach students to solve?

In an article for the \jnl{Bulletin of the Atomic
Scientists}, Bengt Carlson explained how Stanislaw Ulam
invented the technique of supercompression and described it to
Edward Teller.  Carlson upbraids Teller for not
acknowledging Ulam's genius:
\cue{1}
\beginquote
An astonishing
breakthrough was at hand, but where were the signs of [Teller's]
elation?\ft{`How Ulam set the stage', \jnl{Bulletin of the Atomic
Scientists}, July/August 2003.}
\endquote
For Ulam had figured out how to trigger a fusion reaction, or how to
build a hydrogen bomb.

The priority contest of the insane has lasted over 50 years.  Teller
wanted the credit.  Ulam's partisans, such Carlson, award Ulam
the credit.  None say `This instrument of death was not my doing', or
even `I regret my part in its creation.'  Rather, Hans Bethe
remembered
that `The
ideas we had about triggering the H-bomb were all wrong but the
intellectual experience was unforgettable'.\ft{Brian Easlea, {\it
Fathering the Unthinkable: Masculinity, Scientists and the
Nuclear Arms Race\/} (London: Pluto Press, 1983), p. 125.}

Teller was educated in the Hungarian education system,
one of the finest environments for a young
mathematician, where he learnt deep
conceptual understanding.  Ulam, Teller, and Bethe were legendary
problem solvers.  Their child, the hydrogen bomb, shows the
connection between physics and power.  Their eagerness to struggle
with the fascinating problems of building it reflects the disembodied
thinking that physics teaches.

Please, I want no more problem solvers!

\section What I want

I want a world based on mutual aid\ft{Kropotkin, {\it Mutual Aid: A
Factor in Evolution\/}
(1902).  Online at
\url{http://dwardmac.pitzer.edu/Anarchist_Archives/}.}
rather than on power, greed, and the violence to satisfy greed.  I
want my teaching to encourage such a world.  Physics, however, has
grown up as a child of war and to serve power.  By not recognizing this
intimate connection, we supply students who will unwittingly or
unwillingly make the world more miserable.

\section Pure physics and violence

The connection, so clear in building a hydrogen bomb, is also strong
in pure physics.  Many physicists were upset when Congress canceled
the Superconducting Supercollider.  American society, many of us felt,
had abandoned man's ancient quest to comprehend the universe.

Through the US Freedom of Information Act, the physicist Charles
Schwartz obtained the horse's own words, a War Department\ft{As it was
long and accurately known.  It was renamed in 1947 to the
Defense Department.  See Orwell, {\it 1984}.} evaluation
of the Superconducting Supercollider (SSC) just before Reagan approved
its funding:
\cue{2}
\beginquote
  The SSC project will have many spinoffs for the DoD, especially in
  technologies required by the Strategic Defense Initiative, including
  particle beams, information processing, computer control, pulse power
  sources, and high energy accelerators.

  The nuclear weapons community will benefit from the fundamental
  research on the building blocks of atomic matter.  The SSC will
  provide a valuable resource of scientific personnel.  Many of the
  scientists now in the [Department of Energy] nuclear weapons
  laboratory complex received their training while working on particle
  accelerators.\ft{Charles Schwartz, `Political Structuring of the
  Institutions of Science'.  In Laura Nader (editor),
   {\it Naked  Science:  Anthropological Inquiry into Boundaries,
  Power,
   and Knowledge\/} (New York: Routledge, 1996).}
\endquote
Quantum field theory, anyone?

Other physicists were glad about the SSC's demise because its huge
cost sucked money from the smaller, table-top experiments of condensed
matter physics.  The little guy versus Big Science.  Jeff Schmidt
cites a seemingly pure study of matter, `The interaction of
electromagnetic radiation with solid materials':\ft{Jeff Schmidt,
{\it Disciplined
Minds: A Critical Look at Salaried Professionals and the
Soul-Batter\-ing System that Shapes Their Lives\/} (Lanham, Maryland: Rowan$\,$\&$\,$Littlefield, 2000), pp. 73--74.}
\cue{3}
\beginquote
  The objective of the proposed program of theoretical research is an
  increased understanding of the interactions of electromagnetic
  radiation, particularly infrared, with matter.
\endquote
Abstract, pure, and disinterested research?  The War Department
knows wherefore it spends its money.  
Among the benefits it saw:
\cue{4}
\beginquote
  The infrared optical properties of these materials are important to
  \dots [understanding] the interaction of materials subjected to laser
  beams.\ft{Schmidt, p. 74.}
\endquote
Follow the laser beams to the Normandy room session or
onto a front-page {\it New York Times\/} 
article about American snipers in Iraq:\ft{`In Iraq's Murky Battle,
Snipers Offer U.S. a Precision Weapon',
2 Jan 2004, p. A1.} 
\cue{5}
\beginquote
  In an age of satellite-guided bombs dropped at featureless targets
  from 30,000 feet, Army snipers can see the expression on a man's face
  when the bullet hits.
\endquote
Their weapon of choice is `an M-14 rifle equipped with a special optic
sight that has crosshairs and a red aiming dot'.  A sniper
explains the results:
\beginquote
  `I shot one guy in the head, and his head exploded.'
\endquote
Paraphrasing duPont's slogan: 
Better living through optics and GPS.\ft{`Better things for better 
living, through chemistry', introduced in 1939.}

\section What to teach?

Maybe we should not teach physics.  Or we should teach people enough
physics so that they are not intimidated by scientists and they feel
confident enough as citizens to control science.

It is ironic that I cannot talk to you about this topic in person
because I am teaching a physics course.  I am equally caught in this
dilemma and am serving the system.

I mostly teach approximation and order-of-magnitude physics.  A
favorite example of the students is to estimate how many barrels of
oil the United States imports every year.  
\cue{10}
I have used this question
for years, but the invasion of Iraq has made it more relevant.  One
reasonable method is to estimate the number of cars (one per person),
the annual miles driven per car, and convert their product into
gallons and barrels.  How big is a barrel?  Break it into length,
width, and height and multiply.  Eventually you get 3 billion barrels
per year.
$$\myboxit{\vbox{\hsize=5truein%
\def\em#1{{\it #1}}
\parindent=0pt
\centerline{\bf How many barrels of oil does the US import annually?}
\def\m{\u{m}} \def\gal{\u{gal}}
\def\u#1{\,\hbox{#1}}\def\x{\times}
$$3\x10^8\u{people}\x10^4\u{mi/yr}\x{1\u{gal}\over30\u{mi}}
\sim 10^{11}\u{gal}.$$
A barrel has diameter roughly $0.5\m$, so
$$V_{\rm barrel}\sim
1\u{m high}\x0.5\u{m wide}\x0.5\u{m deep}\sim250\,\ell\sim60\gal.$$
\halign{#\hfil\quad&#\hfil\cr
So annual auto consumption:&\em{1.5 billion barrels}\cr
\noalign{\smallskip}
Other transportation, refining inefficiencies:&$\times\,2$.\cr
All other uses:&$\times\,2$.\cr
Fraction imported:&$\times\,0.5$.\cr
\noalign{\medskip}
Imports:&\em{3 billion barrels.}\cr}}}$$
Students learn to break complicated problems into subproblems.  They
learn the power of simple techniques (here, just multiplication) and
enjoy applying physics to the world.  Note how I appeal to the `power'
of this technique.  The connection between physics and power is
difficult to evade!

Physics also guards entry to well-paid, professional careers such as
engineering and medicine.  By eliminating abstruse mathematics, the
order-of-magnitude approach connects physics to the real world and
makes physics less intimidating.  It is humane in a limited way.

Limited, because these estimation skills are what management
consultant firms such as McKinsey look for in their interviews.  One
London firm asks my students: {\it How many hairs are there on a dog?\/}
Microsoft, a convicted corporate criminal, uses similar questions in
its interviews.  People who can answer such questions have developed a
quantitative confidence; they demonstrate that they solve new problems
without fear.  But whose problems?

Nor is order-of-magnitude analysis limited to studying the oiligarchs.
Fermi estimated the size of the first atomic bomb by dropping scraps
of paper and seeing how far backwards the shock wave pushed them.  The
English fluid dynamicist G.~I.~Taylor estimated the size of the same
blast from fireball photographs published in \jnl{Life\/} magazine.
Leading war physicists -- such as Bethe, Feynman, and Fermi -- were
brilliant at approximation.

Again, whose problems am I preparing students to solve?

I mitigate the damage that I do by extending the oil problem.  I ask
students to estimate how much the United States spends on its military
forces that guard Persian Gulf Oil.  
\cue{12}
They rightly guess \$100 billion.
We convert it to the more manageable figure of 35 dollars of military
spending per barrel of oil imported.  This cost is almost exactly the
price of a barrel on the so-called open market.  If the US government
is so concerned to obtain cheap oil for Americans, why not end the
Persian Gulf protection force, use the \$100 billion to buy the oil,
and give it to Americans for free?  I leave this puzzle to you.

How much damage have I undone?  I don't know.  My attempts to go
farther are subtly but strongly resisted.

As a protest against the war on Iraq, I constantly wear an A4-sized
sign:
$$\myboxit{\vbox{\font\tf=pplb7t at 20pt
\myRed \tf
\openup2\jot
\halign{\quad\hfil#\hfil\quad\cr
Oil\rlap{-}\cr
igarchy\cr}\myBlack}}$$
I have worn it since March 2003,
when American and Britain invaded Iraq,
and I wear it while teaching thermodynamics
to the 180 sophomore physics majors.  Students ask about it
in and after class, and I have put my answer on the course website.\ft{\url{http://www.inference.phy.cam.ac.uk/sanjoy/teaching/thermo/}.}  I
explain that as an American citizen, a British citizen, and as a
teacher of physics, a subject that enables war, I feel a triple
obligation to protest.

Fancy formulas, such as the Maxwell relations, are out in this course.
I teach thermodynamics
through estimation, group discussion, and class arguments
over physics.  One problem, for example, asks students to estimate the
thickness of ice formed on lakes during a cold winter; another to
estimate how long a turkey takes to cook.

The conservatives in the department dislike, even despise the
non-traditional approach and say that the course is, in their words,
too political.  They cite my oiligarchy sign and point out
that my course
website condemns exams for ruining learning and making students feel
miserable about themselves.\ft{See Alfie Kohn, {\it Punished by
Rewards: The Trouble with Gold Stars, Incentive Plans, A's, Praise,
and Other Bribes\/} (Boston: Houghton Mifflin, 1993) and 
\url{http://www.alfiekohn.org}.}  The progressives see that students might
learn a lot with the new teaching style (it's the first Cambridge
physics course taught this way, so no one yet knows for sure).  Yet
the progressives, who, like most of Britain, oppose the war, also want
me to remove the sign and the other so-called political aspects.  With
reason on both points, they point out that such prudence will make it
harder for the conservatives to scupper the course and also might make
people take my political message more seriously.

If the conservatives get their way, I would not teach the students to
solve problems skillfully, and the students would have no conception
of the political and social role of physics.  If the progressives get
their way, I would train the students to be skilled problem solvers,
and even enjoy physics, but with no conception of the political and
social role of physics.  Which alternative is worse?

So in class I show the War Department statement about the
Superconducting Supercollider and we discuss the connection between
physics and war.  Can 15 minutes, in one lecture out of hundreds that
students attend in the year, compete against the prestige, funding,
and high salaries in war-financed research (weapons of mass
destruction) or in bond and currency speculation (weapons of financial
mass destruction)?

\section Rule of thumb

To paraphrase John Bright, the anti-tariff and antiwar Victorian
MP:\ft{Bright's quote is in Albert Jay Nock, {\it Our Enemy the State\/}
(1935), Part III, Chapter 2.}
\cue{7}
\beginquote
  The US government, like all powerful governments, has done good
  things, but it has never done a good thing merely because it was a
  good thing.
\endquote
Since I teach approximation
I will speak roughly: The US government
serves the powerful interests that run it, unless a massive popular
movement forces on it an unwelcome policy.  For example, withdrawing
from Vietnam.

Funding for physics increased greatly after World War 2.  Physics
education reform began around the same time and accelerated after
Sputnik.  No popular movement forced the government to fund physics or
education reform.  I therefore wonder: So why the funding?  One answer
is that physicists serve the state.  The state worries when either the
supply or quality of servants falls: {\it I do say, it's\/ {\bf so} hard to
find good help these days.}  To make more and better help is one job
of PER, a field where I have many friends
and colleagues, but a field that I see as dangerous.

\section Cognitive revolution

PER arose from the cognitive revolution in psychology.  Allen Newell,
one of its pioneers, worked at RAND studying the human mind as one
piece of a large `information-processing system' that was the
precursor to NORAD.\ft{Douglas D. Noble, {\it The Classroom Arsenal:
Military Research, Information Technology and Public Education\/}
(London: Falmer Press, 1991), p. 42.}

Another pioneer was Jerome Bruner.  During World War 2 he worked for
the Psychological Warfare Division in the Office of War
Information.\ft{John Rudolph, {\it Scientists in the Classroom\/} (New York:
Palgrave, 2002), p. 98.}  
To education reform he brought the skills of a psychological
warrior.  One flagship NSF-funded curriculum
of the 1950s, PSSC Physics,\ft{Physical Science Study Committee, {\it
PSSC Physics: Teacher's Resource Book and Guide\/} (Bos\-ton: D.~C.~Heath,
1960).}
was designed for a comprehensive assault:
filmstrips, textbooks, readings, and workshops would together overcome
student ignorance and educational conservatism.  In Bruner's words:
\cue{8}
\beginquote
  we [suggested] the analogy to a weapon system -- proposing that the
  teacher, the book, the laboratory, the teaching machine, the film, and
  the organization of the craft might serve to form a balanced teaching
  system.\ft{Rudolph, p. 99.}
\endquote
As historian John Rudolph concludes:
\beginquote
  The modern Cold War weapon system was, in the minds of all these
  reformers, the epitome of rational instrumentation -- a powerful
  model to be emulated in seeking solutions to educational
  problems.\ft{Rudolph, p. 99.}
\endquote
I conceive of education as growth in exploratory powers, curiosity,
and confidence, described so well by Hassler
Whitney\ft{Hassler Whitney, `Coming alive in school math and beyond',
\jnl{Educational Studies in Mathematics\/} 18(3):229--242 (1987).},
Louis Benezet,\ft{L.~P.~Benezet, `The Teaching of Arithmetic I, II, III: The
Story of an Experiment', \jnl{Journal of the National Education
Association\/} 24:241--244, 301--303 (1935); 
25:7--8 (1936).  The articles 
are online at
\url{http://www.inference.phy.cam.ac.uk/sanjoy/benezet/}.}
and John Holt.\ft{John Holt, {\it How Children Learn\/} (Penguin,
1983); and {\it How Children Fail\/} (Penguin, 1983).}  However, in
the Sputnik-era reforms, the student mind was a target to be measured,
mapped, re-engineered, and overcome; and cognitive psychology provided
the ammunition.

\section PER As Ideological Facelift

But why PER?

In 1958, only 18\%\ of Americans thought that their government was `run
by a few big interests looking out for themselves.'\ft{Earl Babbie,
{\it You Can Make a Difference: The Heroic Potential Within Us
All\/} (New York: St. Martin's Press, 1985), Chapter 3.
Online at
\url{http://www1.chapman.edu/wilkinson/socsci/sociology/Faculty/Babbie/YCMAD/}.}
But after the Vietnam war, many institutions of American society took
a drubbing.  The President resigned rather than let us know what other
crimes he was involved in.  The secret police (the FBI) was exposed
spying on American citizens.  The CIA was exposed overthrowing the
governments of Guatemala, Iran, and Chile to install torturing
kleptocrats friendly to American corporations.

In 1980 the 18\%\ who thought the government was run `by a few big
interests' had become 77\%.  Our ruling institutions needed an
ideological facelift.  The presidency of Jimmy Carter and its concern
for human rights came at the right time.  As rhetoric about human
rights flowed liberally, so did torture equipment for the Shah and
tanks and planes for Indonesia to slaughter in East Timor.

Science suffered a similar sag in prestige.  Dow Chemical became
notorious for manufacturing Agent Orange and napalm.  In {\it Silent
Spring}, Rachel Carson exposed how corporations poisoned the
environment with the products of science.  Scientists, the servants of
the discredited state, lost public trust, and science itself acquired
a bad odor.

Where would the state get more servants?  Would the rest of society
trust these servants?  Science needed many ideological nips, tucks,
and lifts.

Developing at the right time from the cognitive revolution, PER
arose to make a kindler, gentler physics.  Students would work in
groups.  Teachers would find out what students knew and teach
accordingly.  Students would do hands-on experiments and inquiry
lessons; they would learn how scientists investigate the world, and
they would be inducted into the so-called scientific method and
epistemology.  Students would become happy, creative problem solvers
and would learn that true science, unlike the authoritarian science
represented through the old teaching style, is a force for reason and
good.

But this new science is not so new.  For example, the Views about
Science Survey (VASS) claims to measure how `expert' a student's views
about science are.  No question mentions the connection between
science and war, because such a connection is outside the expert
(i.e. scientist) view of science.  Yet in 1986, of recently graduated
physics bachelors working in science or engineering, one-half
worked on war projects.\ft{Charles Schwartz, `Social responsibility in
  physics', \jnl{Social Responsibility}, vol 2(1).  Online
  at \url{http://socrates.berkeley.edu/~schwrtz/PhysBklt.html}.}
If physics is what physicists do, does your PER-based physics course
teach this connection?  Or do you strictly teach problem solving?

\section 1984 vs Brave New World

Imperial rulers can choose from two strategies to coerce the people to
work for the elite: {\it 1984}, which uses physical force; or {\it
Brave New World}, which uses psychology.  PER is a timely hybrid: It
employs advanced social science to further advanced physical science.

Physical science dominates in the 1984 approach.  We prefer it for the
browner peoples of the far empire: Vietnam, Cambodia, Indonesia, Iran,
Iraq, and for the darkest at home.  Thanks to the war `on'
terror, the less dark at home also feel the lash.  But normally
the richer of the home population,
to amuse themselves into passivity,
get a brave new world of Fox
`News', Super Bowl, Big Brother, cheap DVD recorders, and
Hollywood movies and celebrities.  As
Aldous Huxley warned:
\cue{9}
\beginquote
  The most important Manhattan Projects of the future will be vast
  government-sponsored enquiries into what the politicians and the
  participating scientists will call `the problem of happiness' - in
  other words, the problem of making people love their
  servitude.\ft{Huxley's Foreword to {\it Brave New World} (1946).}
\endquote
One example is PER, which wraps service to power in a humane shell.

\section War budget

In the fiscal year 2002, we -- via our government -- spent almost
\$600 billion on death.
\cue{10}
That figure includes the nominal war
budget plus its many disguises including veterans pensions, foreign
`aid' to buy our weapons, nuclear-weapons
programs in the Department of Energy, and interest on the national
debt due to past war spending.\ft{Robert Higgs, `The Defense Budget Is
Bigger Than You Think',
\url{http://www.independent.org/tii/news/031222Higgs.html}.}

\topinsert
\vbox{{\openup0\jot \def\1{\noalign{\smallskip\hrule \vskip1pt
\hrule\smallskip}}
\centerline{\it Table 1.  US military spending in Fiscal Year
2002.}
\smallskip
\halign{\hfil#\qquad&#\hfil\cr
\1
\$billions&{\it Where\/} (from Higgs, `Defense budget')\cr
\noalign{\smallskip}
344.4&War Department itself\cr
138.7&Portion of interest on national debt due to past wars\cr
50.9&Department of Veterans Administration\cr
18.5&Defense-related parts of the Department
of Energy budget\cr
17.6&State Department and international assistance
programs related to 'defense'\cr
17.5&Agencies later incorporated into 
Department of Homeland Security\cr
8.5&Other homeland security\cr
\noalign{\smallskip}
\it 596.1&\it Total\cr
\1
}}}
\endinsert

Killing consumed about one-half of the
government budget even before we invaded Iraq.  In 1967 Martin Luther
King warned:
\beginquote
A nation that continues year after year to spend more
money on military defense than on programs of social uplift is
approaching spiritual death.\ft{`Declaration of Independence from the
war in Vietnam', 4 April 1967, Riverside Church, New York, NY.}
\endquote
One year later, he was shot dead, probably by the US
government.\ft{See Bill Pepper, {\it An Act of State: The Execution of
Martin Luther King\/} (London: Verso, 2003).}

\section Hippocrates

Modern science and the modern, absolute state arose together in the
17th century.  We got the heliocentric theory and Louis XIV, the Sun
King, who proclaimed that `The state is I'.\ft{Lewis Mumford, {\it The
Myth of the Machine: The Pentagon of Power\/} (London: Secker and
Warburg, 1970), especially Chapters 2 and 7.}  Both science and the
state worship Law with a capital L.  Rulers, whether the soft American
or hard Russian variety, depend on science for ideological and
physical power.

As long as there is an elite, physics will serve it and cement its
rule.  When we develop a society without an elite, we can reconsider
whether to teach and fund physics.  Until then should we follow
Hippocrates?  He advised doctors as their first commandment: {\it
First do no harm.}  A step in stopping an evil is refusal to
cooperate.\ft{\'Etienne de la Bo\'etie,
{\it The Politics of Obedience: The Discourse of Voluntary Servitude\/}
(New York: Free Life Editions, 1975 [1552--53]).  Online with an
introduction by Murray Rothbard at
\url{http://www.mises.org/rothbard/boetie.pdf}.}
Don't teach physics.  By teaching physics, traditionally
or in the reformed sense of PER, we train more servants of the state
and we legitimize science.  As we thereby arm the state against
ourselves, we move farther from a peaceful, cooperative, and enjoyable
world.

\end